\documentclass[12pt]{article}

\usepackage{amsmath,amssymb,amsthm}
\usepackage{cite}
\usepackage{graphicx}
\usepackage{caption2}

\newtheorem*{theorema}{Janet-Cartan theorem}

\begin{document}

\title{Embeddings of the black holes in a flat space}

\author{
A.A.~Sheykin\thanks{E-mail: anton.shejkin@gmail.com},
D.A.~Grad\thanks{E-mail: nirowulf239@gmail.com},
S.A.~Paston\thanks{E-mail: paston@pobox.spbu.ru}\\
{\it Saint Petersburg State University, St.-Petersburg, Russia}
}
\date{\vskip 15mm}
\maketitle

\begin{abstract}
We study the explicit embeddings of static black holes. We obtain two new minimal embeddings of the Schwarzchild-de Sitter metric which smoothly cover both horizons of this metric. The lines of time for these embeddings are more complicated than hyperbolas.
Also we shortly discuss the possibility of using non-hyperbolic embeddings for calculation of the black hole Hawking temperature in the Deser and Levin approach.
\end{abstract}

\newpage

\section{Introduction}
The problem of connection between internal and external geometry of a surface embedded in some ambient space has a long story. First results related to embeddings were obtained by Gauss and Riemann in the middle of the XIX century. In 1916 Janet and Cartan proved the existence of an isometric embedding of a Riemannian manifold (the Janet-Cartan theorem was generalized on pseudo-Riemannian case by Friedman in 1961) \cite{goenner}:

\begin{theorema}
Any $n$-dimensional Riemannian manifold $\mathcal M_n$ with an analytic metric can be locally and isometrically embedded in an $N$-dimensional Euclidean space $E_N$ where $N=n(n+1)/2$.
\end{theorema}

The metric on this manifold becomes induced and can be expressed in terms of an embedding function:
\begin{align}
	g_{\mu\nu}(x) = \partial_\mu y^a(x) \partial_\nu y^b(x) \eta_{ab},\label{met}
\end{align}
where $y^a(x^{\mu})$ is the embedding function and $\eta_{ab}$ is the ambient Minkowski space metric.

The number $p=N-n=n(n-1)/2$ is called the embedding class. It is easy to see that for an arbitrary 4-dimensional manifold $p=6$. In the case of a manifold with some symmetries the embedding class might decrease; e.g., for constant curvature spaces $p=1$, and for spherically symmetric spaces $p=2$ \cite{schmutzer}.

The exploitation of embeddings in the general relativity began shortly after its appearing. In particular, the first embedding of the Schwarzchild metric was constructed by Kasner in 1921, just five years after discovering the metric itself.

The embeddings can be of use in general relativity for many purposes. Until 70's the embeddings had been used mainly for classification of exact solutions (as the embedding class of a given metric is invariant) \cite{schmutzer} of the Einstein equations and for examination of geometrical structure of metrics -- for example, the famous Kruskal coordinates were possibly obtained using the embedding method \cite{frons}.

As we know, 70's were the years of the string theory rising. From the geometrical point of view it is a theory of a two-dimensional curved space embedded in a Minkowski space. In 1975 Regge and Teitelboim developed a string-inspired approach to general relativity  \cite{regge}, in which our four-dimensional spacetime is considered as a surface in the ten-dimensional Minkowsky space. In this case the embedding function becomes a dynamical variable, and the Einstein equations are replaced by the Regge-Teitelboim (RT) equations:
\begin{align}
	\partial_{\mu}\Bigl( \sqrt{-g} (G^{\mu\nu} -\varkappa T^{\mu\nu} ) \partial_{\nu} y^a\Bigr) =	0.\label{RT}
\end{align}
The natural appearance of the Minkowski space in this formulation might ease the quantization -- in particular, one can use the ambient space for defining the causality, and the timelike direction of this space can play the role of time in quantization. Several variants of a canonical formalism for such a theory were constructed in papers \cite{statja18,statja24} and \cite{davkar}; in the latter work was also obtained the Wheeler-de Witt equation for the Friedmann universe in terms of the embedding function.

To avoid the nesessarity of introducing unobservable coordinates on the embedded manifold, the RT theory was reformulated in \cite{statja25} as a field theory in the flat ambient space -- if one defines a set of scalar fields in the ambient space, then some surface in this space can be defined as a surface of constant values of these fields.

 Unfortunately, the Regge-Teitelboim theory has it own difficulties. Among them should be mentioned thr fact that the RT equations are more general than the Einstein equations, so in the RT theory the so-called "extra solutions" exist \cite{deser}. It was shown in the work \cite{statja26} that with certain cosmological assumptions the influence of these solutions on the observables at the present time is negligibly small.
To solve the problem of extra solutions, L.D. Faddeev proposed the variant of an embedding theory in the work \cite{faddeev}, in which the dynamical variable is not an embedding function, but its derivative, so no extra differentiation in the field equations appear.  Also there were some attempts to interpret these extra solutions as a possible source of the dark matter \cite{davids01, estabrook}.

\section{Explicit embeddings}
In the studying of the geometrical properties of manifolds and structure (\ref{RT}) the problem of constructing the explicit embedding of a given metric, formally equivalent to the problem of solving the system (\ref{met})  for $y^a(x)$  arises. Since in general case it is a nonlinear system of PDE's, no regular algorithm of constructing an embedding for arbitrary metric is known. However for some physically interesting spacetimes due to their high symmetry one can find explicit embeddings. Particular examples of spacetimes with such property are the Friedmann universe, plane-fronted gravitational waves and static black holes. One can find many explicit embeddings of these and other spacetimes in the reviews \cite{rosen65, collinson68}.

In spite of several attempts to systematically construct explicit embeddings \cite{fujitani,collinson68}, until recent time this construction had been performed rather intuitively. Thus two global embeddings of the Schwarzchild metric in the six-dimensional Minkowsky space remained unknown and were recently obtained in \cite{statja27} using the method based on the group representation theory. In the same manner were firstly obtained three minimal embeddings of Reissner-Nordstr$\ddot{\text{o}}$m metric \cite{statja30}, which smoothly cover both horizons of this metric. As we think, all these embeddings remained unknown because they realize the invariance of the metric under the shifts of $t$ through not only (hyper)rotations in the ambient space, but through rotations conjugated by translations. The translations were used for the first time in \cite{davidson} and later in \cite{blashke2010}. Note that the problem of representation of time shifts as some Poincar$\acute{\text{e}}$ transformations was studied also in  \cite{letaw81} but for another purpose -- in order to obtain all possible stationary trajectories of motion in a Minkowsky space.  Nevertheless, the results of this analysis mainly correspond to the results given in \cite{statja27}.

Using the method proposed in  \cite{statja27} one can obtain an explicit embedding of any metric with enough symmetry (e.g. a static black hole), but this embedding turns out to be not nesessarily global, i.e. smooth for all values of radius $r$. If we are interested only in global embeddings in a six-dimensional ambient space, we can construct the generalization of 2 of 6 Schwarzchild metric embeddings listed in \cite{statja27} to the case of the existing positive $\Lambda$-term (SdS):
\begin{align}
	ds^2=\left(1-\frac{2m}{r}-\frac{\Lambda r^2}{3}\right)dt^2-\dfrac{dr^2}{\left(1-\dfrac{2m}{r}-\dfrac{\Lambda r^2}{3}\right)} - r^2d\Omega^2,
\end{align}
where $m$ is the black hole mass, $\Lambda$ is the cosmological constant. In the physically interesting case $0\leq \Lambda \leq m^{-2}/9$.

The first embedding is a spiral embedding in the Minkowsky space with the signature $(+-----)$:
\begin{align}
      \begin{array}{lcl}
\displaystyle y^0=u, & \qquad & y^3 = r \cos \theta,      \\[0.8em]
\displaystyle y^1=\frac{f(r)}{\alpha}\,\sin\left(\alpha{u}-\psi(r)\right), & \qquad &  y^4 = r \sin \theta \cos \phi,\\[0.8em]
\displaystyle y^2=\frac{f(r)}{\alpha}\,\cos\left(\alpha{u}-\psi(r)\right),  & \qquad &  y^5 = r \sin \theta \sin \phi,\\
      \end{array}\label{emb}
\end{align}
where \begin{gather}
f(r)=\sqrt{\frac{2m+\Lambda r^3/3}{r}},\qquad \alpha\geq \frac{1}{(6\sqrt{3}m)}\label{res},\\
	\psi(r)=\pm\int\! {dr}\sqrt{\alpha^2+f'^2(1-1/f^2)}\label{psi}
\end{gather}
and the retarded time $u$ is related to the Schwarzchild time as follows:
\begin{align}
	u=t+\frac{1}{\alpha}\int\! dr\, \frac{f^2\psi'}{1-f^2}.
\end{align}

The second one is a cubic embedding in the same space:
\begin{align}
	\begin{array}{cl}
	&\displaystyle y^{0,1}= \frac{\xi^2}{6}u^3 +\frac{h(r)\pm 1}{2}\,u + \chi(r),\\
&\displaystyle y^{2} =\frac{\xi}{2}u^2+\frac{1}{2\xi}h(r)
	\end{array}\label{emb1}
\end{align}
and $y^3$, $y^4$, $y^5$ are the same that in (\ref{emb}). Here
 \begin{gather}
h(r)=1-\frac{2m}{r}-\frac{\Lambda r^2}{3},\qquad \xi\ge\frac{\sqrt{27}}{64m},\label{res1}\\
	\chi(r)=\pm\int\! {dr}\,\sqrt{1-h-\frac{h h'^2}{4\xi^2}}\label{chi}
\end{gather}
and
\begin{align}
	u=t+\int\! dr\, \frac{\chi'}{h}.
\end{align}

One can prove that with the restrictions (\ref{res}), (\ref{res1}) on the parameters $\alpha$ and $\xi$ the radicands in (\ref{psi}) and (\ref{chi}) are non-negative for all values of $r$, and these embeddings are global, i.e. smooth for all $r$ with a proper selection of sign in  (\ref{psi}) and (\ref{chi}) for different values of $r$.  As for all embeddings using the translations in the ambient space, these two embeddings do not cover all regions corresponding to a maximal analytical extension of the metric, but only one copy of regions inside, between and out of the horizons (in the case when the black hole horizon $r=r_b$ and the cosmological horizon $r=r_c$ exist) ; these regions are bordered by the thick line on fig.1.  Thus the embedding surface is geodesically incomplete: there are timelike geodesics which go to infinity in a finite proper time.
\begin{figure}[htbp]
\begin{center}
\includegraphics[width=30em]{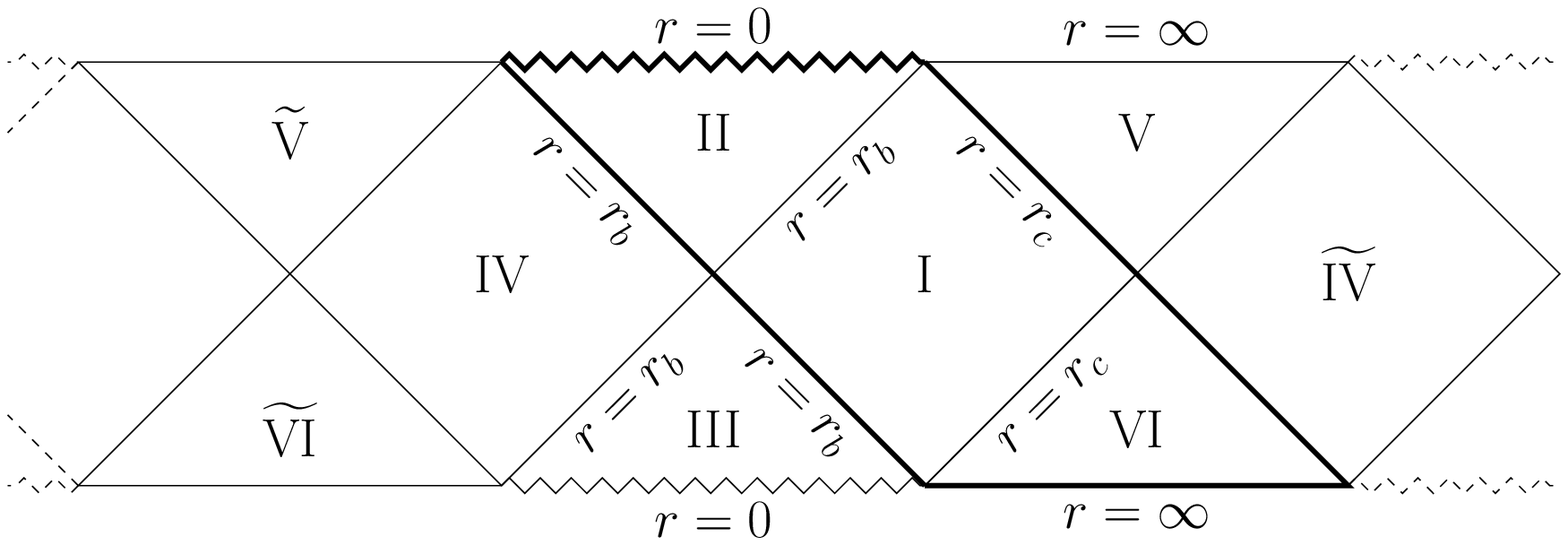}%
\caption{
Penrose diagram for a maximal analytical extension of the SdS metric.
I -- our universe; II -- black hole, III -- white hole, IV, $\widetilde{\text{IV}}$ -- parallel universes, V, VI,
$\widetilde{\text{V}}$, $\widetilde{\text{VI}}$ -- the regions with $r>r_c$.
A part covered by the embeddings
(\protect\ref{emb}) and (\protect\ref{emb1})
is bordered by the thick line.
}
\end{center}
\end{figure}

It is easy to see that these embeddinds admit the limit  $\Lambda \to 0$ in which they turn into the spiral and cubic embeddings \cite{statja29} of the Schwarzchild metric respectively. It is possible due to the fact that in this limit the asymptotics of the embedding in $r\to 0$ remains unchanged. Note that in the case of global embeddings of the RN metric given in \cite{statja30} the situation is opposite: these embeddings turn out to be singular at $q \to 0$ limit because of a change in such an asymptotics. Also it is impossible to embed the S-AdS metric globally in the six-dimensional Minkowski space because of the properties of its asymptotics.

Of course, among various types of black holes those which have a nonzero angular moment -- Kerr and Kerr-Newman black holes -- are of most interest. However, the construction of an explicit embedding of such a spacetime is a nontrivial problem that still remains unresolved. The only known embedding of the Kerr metric is the Kuzeev embedding \cite{kuzeev} written in an implicit form (there is a PDE on two components of the embedding function). It is a local embedding in (3+6)-dimensional Minkowski space which do not contain translations in the ambient space. Adding such translations possibly might help to lower the embedding class or to obtain an explicit embedding of the Kerr metric, but this question requires additional investigation.

\section{Thermodynamics of black holes}
The interest in the explicit embeddings of black holes has significantly increased after the paper \cite{deserlev99} in which these embeddings were used for study of thermodynamical properties of black holes. In this work Deser and Levin discovered the mapping between the parameters of the Hawking radiation of the black hole and Unruh radiation detected by the observer which is moving on the surface of the	 embedding corresponding to this black hole. This correspondence was checked for Fronsdal embedding of the Schwarzschild metric and for similar (hyperbolic) embeddings of RN and S-(A)dS metrics.

By the time of the publication of \cite{deserlev99} there was only one type of embedding which smoothly covers the horizon of the black hole, namely the hyperbolic one, whereas now three additional types of such embedding is known \cite{davidson,statja27}. Thus it is interesting to check the mapping between the Hawking and Unruh radiation when these types of embeddings are used.

Since the Unruh temperature is usually related to the acceleration of a test particle, we can calculate first	 the acceleration of a particle moving on the embedding surface in ambient space. Consider the three Schwarzchild metric embeddings which contain translations. The corresponding "six-dimensional" accelerations $a_6$ of the particle which is rest at the distance $r$ have the form
\begin{align}\label{77}
\dfrac{1}{3\sqrt{6 m r}\left(1-\dfrac{2m}{r}\right)},\qquad
\dfrac{\beta}{\left(1-\dfrac{2m}{r}\right)} \sqrt{1-\dfrac{2m}{r}+\hat\gamma^2},\qquad
\dfrac{\xi}{\left(1-\dfrac{2m}{r}\right)}
\end{align}
for spiral (asymptotically flat), Davidson-Paz and cubic embeddings respestively. Here $\beta$, $\hat\gamma$ and $\xi$ are the parameters of the corresponding embeddings, see \cite{statja29}. For comparison the  "six-dimensional" acceleration $a_6$ for the Fronsdal (hyperbolic) embedding has the form
\begin{align}
a_6=\frac{1}{4m\sqrt{1-\dfrac{2m}{r}}}.
\end{align}

In the Fronsdal embedding the trajectory of the particle in the ambient space is a hyperbola, which corresponds to the uniformly accelerated motion. In this case the Unruh radiation with the thermal spectrum and temperature $T=a/(2\pi)$ is detected. It is easy to 	see that in this case Tolman's law $T\sqrt{g_{00}}=\text{const}\equiv T_0$  is satisfied and $T_0=1/(8\pi m)$ matches the Hawking temperature.

For new embeddings which involve translations in the ambient space the trajectories are no longer hyperbolae, although they remain stationary (they are studied in \cite{letaw81}) so for such embeddings the Unruh radiation is modified and is no longer thermal \cite{letaw81,abdolrahimi}. Even if the spectrum of this radiation is approximately thermal and it is possible to use the same relation between the acceleration and the temperature (see \cite{barbado}), one can see from (\ref{77}) that in this case the Tolman's law is violated and the mapping between Hawking and Unruh radiation is absent.

Thus the mapping discovered in \cite{deserlev99} turns out to be non-universal; it holds only for hyperbolic embeddings. The reason of this uniqueness remains unknown. Possibly it is due the fact that embeddings with translations cannot cover all the regions corresponding to maximal analytical extension of the metric, whereas the Fronsdal-like embedding covers the whole manifold. This question requires additional study.

\vskip 0.5em
{\bf Acknowledgments}.
The work of A.A.S. and S.A.P. was partially supported by the Saint Petersburg State University grant N~11.38.660.2013.


\begin{thebibliography}{99}

\bibitem{goenner}
H.~Goenner,
 Local isometric embedding of Riemannian manifolds and Einstein's
  theory of gravitation,
 in {\em General Relativity and Gravitation: one hundred years after
  the birth of Albert Einstein}, edited by A.~Held, vol. 1, pp.
  441--468, Plenum Press, 1980.

\bibitem{schmutzer}
H.~Stephani~et al.,
 {\em Exact Solutions of Einstein's Field Equations, 2nd ed.}
  (Cambridge University Press, 2003).

\bibitem{frons}
C.~Fronsdal,
 Phys. Rev. {\bf 116}, 778 (1959).

\bibitem{regge}
T.~Regge and C.~Teitelboim,
 General relativity \`a la string: a progress report,
 in {\em Proceedings of the First Marcel Grossmann Meeting, Trieste,
  Italy, 1975}, edited by R.~Ruffini, pp. 77--88, 1977.

\bibitem{statja18}
S.~A. Paston and V.~A. Franke,
 Theor. Math. Phys. {\bf 153}, 1582 (2007), [arXiv:0711.0576].

\bibitem{statja24}
S.~A. Paston and A.~N. Semenova,
 Int. J. Theor. Phys. {\bf 49}, 2648 (2010), [arXiv:1003.0172].

\bibitem{davkar}
D.~Karasik and A.~Davidson,
 Phys. Rev. D {\bf 67}, 064012 (2003), [arXiv:gr-qc/0207061].

\bibitem{statja25}
S.~A. Paston,
 Theor. Math. Phys. {\bf 169}, 1600 (2011), [arXiv:1111.1104].

\bibitem{deser}
S.~Deser, F.~A.~E. Pirani, and D.~C. Robinson,
 Phys. Rev. D {\bf 14}, 3301 (1976).

\bibitem{statja26}
S.~A. Paston and A.~A. Sheykin,
 Int. J. Mod. Phys. D {\bf 21}, 1250043 (2012), [arXiv:1106.5212].

\bibitem{faddeev}
L.~D. Faddeev,
 Theor. Math. Phys. {\bf 166}, 279 (2011), [arXiv:0911.0282].

\bibitem{davids01}
A.~Davidson, D.~Karasik, and Y.~Lederer, \textit{Cold Dark Matter from Dark Energy}
 (2001), [arXiv:gr-qc/0111107].

\bibitem{estabrook}
F.~Estabrook,
 SIGMA {\bf 9}, 012 (2013), [arXiv:1206:5229].

\bibitem{rosen65}
J.~Rosen,
 Rev. Mod. Phys. {\bf 37}, 204 (1965).

\bibitem{collinson68}
C.~D. Collinson,
 J. Math. Phys. {\bf 9}, 403 (1968).

\bibitem{fujitani}
T.~Fujitani, M.~Ikeda, and M.~Matsumoto,
 J. Math. Kyoto Univ. {\bf 1}, 43 (1961).

\bibitem{statja27}
S.~A. Paston and A.~A. Sheykin,
 Class. Quant. Grav. {\bf 29}, 095022 (2012), [arXiv:1202.1204].

\bibitem{statja30}
S.~A. Paston and A.~A. Sheykin, \textit{Global embedding of the Reissner-Nordstrom metric in the flat ambient space}
 (2013), [arXiv:1304:6550].

\bibitem{davidson}
A.~Davidson and U.~Paz,
 Found. Phys. {\bf 30}, 785 (2000).

\bibitem{blashke2010}
D.~N. Blaschke and H.~Steinacker,
 Class. Quantum. Grav. {\bf 27}, 185020 (2010), [arXiv:1005.0499].

\bibitem{letaw81}
J.~Letaw,
 Phys. Rev. D {\bf 23}, 1709 (1981).

\bibitem{statja29}
S.~A. Paston and A.~A. Sheykin,
 Theor. Math. Phys {\bf 175}, 806 (2013), [arXiv:1306.4826].

\bibitem{kuzeev}
R.~R. Kuzeev,
 Gravitatsiya i Teoriya Otnositelnosti {\bf 18}, 75 (1981; in
  Russian).

\bibitem{deserlev99}
S.~Deser and O.~Levin,
 Phys. Rev. D {\bf 59}, 064004 (1999), [arXiv:hep-th/9809159].

\bibitem{abdolrahimi}
S.~Abdolrahimi, \textit{Velocity Effects on an Accelerated Unruh-DeWitt Detectors}
 (2013), [arXiv:1304.4237].

\bibitem{barbado}
L.~C. Barbado and M.~Visser,
 Phys. Rev. D {\bf 86}, 084011 (2012), [arXiv:1207.5525].

\end{thebibliography}
\end{document}